\documentclass[twocolumn,showpacs,preprintnumbers,amsmath,amssymb]{revtex4}

\usepackage{amsmath,amssymb,graphicx}
\usepackage{graphicx}
\usepackage{dcolumn}
\usepackage{bm}
\usepackage{epsfig}
\usepackage{here}
\usepackage{float}
\usepackage{amsmath}
\usepackage[usenames]{color}

\newcommand{\bea}{\begin{eqnarray}}
\newcommand{\eea}{\end{eqnarray}}

\begin{document}
\draft


\title{Observational constraints on a unified dark matter and dark energy
model \\ based on generalized Chaplygin gas}
\author{Chan-Gyung Park${}^{1}$, Jai-chan Hwang${}^{1}$, Jaehong Park${}^{1}$,
        and Hyerim Noh${}^{2}$}
\address{${}^{1}$Department of Astronomy and Atmospheric Sciences,
                 Kyungpook National University, Taegu, Korea \\
         ${}^{2}$Korea Astronomy and Space Science Institute,
                 Daejon, Korea}

\date{\today}

\begin{abstract}
We study a generalized version of Chaplygin gas as unified model of
dark matter and dark energy. Using realistic theoretical models and
the currently available observational data from the age of the
universe, the expansion history based on the type Ia supernovae, the
matter power spectrum, the cosmic microwave background radiation
anisotropy power spectra, and the perturbation growth factor we put
the unified model under observational test. As the model has only
two free parameters in the {\it flat} Friedmann background
[$\Lambda$CDM (cold dark matter) model has only one free parameter]
we show that the model is already tightly constrained by currently
available observations. The only parameter space extremely close to
the $\Lambda$CDM model is allowed in this unified model.
\end{abstract}

\noindent 

\maketitle

%
%
\section{Introduction}

A unified dark matter and dark energy model motivated by the
(generalized) Chaplygin gas has been introduced in the literature
\cite{Kamenshchik-etal-2001,Bilic-etal-2002,Bento-etal-2002}, and
has been extensively investigated in the cosmological studies
\cite{Bean-Dore-2003,Bento-etal-2003,Carturan-Finelli-2003,Avelino-etal-2003,
Amendola-etal-2003,Bento-etal-2004,Silva-Bertolami-2003,Bertolami-etal-2004,
Cunha-etal-2004,Zhu-2004,Dev-etal-2003,
Gorini-etal-2003,Makler-etal-2003a,
Makler-etal-2003b,Alam-etal-2003,Colistete-etal-2004,Dev-etal-2004,
Multamaki-etal-2004,Sandvik-etal-2004,Beca-etal-2003,Alcaniz-Lima-2005,
Colistete-Fabris-2005,Bento-etal-2005,Biesiada-etal-2005,Gong-2005,
Lazkoz-etal-2005,Bertolami-Silva-2006,Zhang-Zhu-2006,Zhang-etal-2006,
Wu-Yu-2007a,Wu-Yu-2007b,Guo-Zhang-2007,Barreiro-etal-2008,Gorini-etal-2008,
Fabris-etal-2008,Li-etal-2009,Urakawa-Kobayashi-2009}. Many
constraints on the Chaplygin gas model parameters have been placed
(predicted) based on current (upcoming) astronomical observations
such as type Ia supernovae (SNIa)
\cite{Bean-Dore-2003,Amendola-etal-2003,Avelino-etal-2003,Makler-etal-2003a,
Makler-etal-2003b,Beca-etal-2003,Bertolami-etal-2004,Colistete-etal-2004,
Dev-etal-2004,Cunha-etal-2004,Zhu-2004,Colistete-Fabris-2005,Bento-etal-2005,
Biesiada-etal-2005,Gong-2005,Zhang-etal-2006,Guo-Zhang-2007,Alcaniz-Lima-2005,
Lazkoz-etal-2005,Wu-Yu-2007b,Alam-etal-2003,Silva-Bertolami-2003},
large-scale structure
\cite{Bean-Dore-2003,Avelino-etal-2003,Sandvik-etal-2004,Beca-etal-2003,
Multamaki-etal-2004,Zhang-etal-2006,Gorini-etal-2008,Fabris-etal-2008,
Urakawa-Kobayashi-2009}, cosmic microwave background radiation (CMB)
\cite{Bean-Dore-2003,Bento-etal-2003,Carturan-Finelli-2003,Amendola-etal-2003,
Zhang-etal-2006,Barreiro-etal-2008,Urakawa-Kobayashi-2009},
gravitational lensing
\cite{Dev-etal-2003,Dev-etal-2004,Silva-Bertolami-2003,Makler-etal-2003b},
gamma-ray bursts \cite{Bertolami-Silva-2006}, X-ray luminosity of
galaxy clusters \cite{Makler-etal-2003b,Cunha-etal-2004,Zhu-2004},
look-back time-redshift data \cite{Gong-2005,Li-etal-2009}, angular
size-redshift data \cite{Alcaniz-Lima-2005}, Hubble
parameter-redshift data \cite{Wu-Yu-2007a,Wu-Yu-2007b},
Fanaroff-Riley type IIb radio galaxies
\cite{Makler-etal-2003b,Zhu-2004}, and so on.

It is known that such a unified model is constrained by observations
so that parameter space close to the conventional $\Lambda$CDM
models is allowed. But current status of the model in parameter
space different from close-to-$\Lambda$CDM has been unclear. The
distance-redshift relation, the age of the universe, abundances of
light elements, the large-scale matter density and velocity power
spectrum, the CMB temperature and polarization anisotropy power
spectra, and the perturbation growth factor can be regarded as the
main pillars of modern cosmology where theories meet with
observations. Here, we investigate the unified model by comparing
realistic theoretical predictions with currently available
observations. Our conclusions can be found in section
\ref{sec:Discussion}.

%
%
\section{Generalized Chaplygin gas}

We introduce a fluid $X$ with an equation of state \bea
   \tilde p_X = - A \tilde \mu_X^{-\alpha},
   \label{GCG-EOM}
\eea where $\tilde p$ and $\tilde \mu$ are the pressure and the
energy density, and $A$ and $\alpha$ are constants; tildes indicate
covariant quantities. The Chaplygin gas is a case with $\alpha = 1$.
It is interesting to note that any fluid with barotropic equation of
state, $\tilde p_X = \tilde p_X (\tilde \mu_X)$, has an exact scalar
field theory counterpart with a tachyonic kinetic term given by an
action (we set $8 \pi G \equiv 1 \equiv c$) \bea
   S = \int d^4 x \sqrt{-\tilde g} \left[
       {1 \over 2} \tilde R + \tilde p_X (\tilde X) \right], \quad
       \tilde X \propto e^{\int {2 d \tilde p_X \over
       \tilde \mu_X + \tilde p_X}},
   \label{action}
\eea where $\tilde X \equiv {1 \over 2} \tilde \phi^{,c} \tilde
\phi_{,c}$; see Appendix A for a proof. For a fluid in Eq.\
(\ref{GCG-EOM}) we have \bea
   \tilde p_X \equiv - A^{1 \over 1+\alpha} \left[ 1
       - \left( {\tilde X /B} \right)^{1+\alpha \over 2 \alpha}
       \right]^{\alpha \over 1 + \alpha},
   \label{action-GCG}
\eea where $B$ is a constant. We call the fluid or field based on
Eqs.\ (\ref{GCG-EOM}) and (\ref{action-GCG}), a generalized
Chaplygin gas (GCG).

%
%
To the background order in Friedmann world model the energy
conservation equation for GCG component gives \bea
   & & \mu_X
       = \left( A
       + { \mu_{X0}^{1 + \alpha} - A
       \over a^{3( 1 + \alpha)} }
       \right)^{1 \over 1 + \alpha}, \quad
       p_X = - A \mu_X^{-\alpha}
   \nonumber \\
   & &
       w_X \equiv {p_X \over \mu_X}
       = - \left( 1 + { \mu_{X0}^{1 + \alpha}/A - 1
       \over a^{3( 1 + \alpha)} }
        \right)^{-1},
   \nonumber \\
   & & c_X^2 \equiv {\dot p_X \over \dot \mu_X}
       = - \alpha w_X,
   \label{mu-GCG}
\eea where $a$ is the scale factor and $\mu_{X0}^{1+\alpha} > A$ (we
set $a_0 \equiv 1$ at the present epoch). The equation of state
parameter $w_X$ is negligible in the early era thus the GCG
potentially acts as a dark matter, and approaches $w_{X0} = -
A/\mu_{X0}^{1+\alpha} (> -1)$ in recent past thus it potentially
acts as a dark energy depending on suitable choice of parameters.

We consider a {\it flat} background with baryon, radiation (photons
and neutrinos), and the GCG. At the present epoch the Friedmann
equation gives \bea
   \Omega_{X0} = 1 - \Omega_{b0} - \Omega_{r0},
   \label{Omega_X0}
\eea where $\Omega_i \equiv \mu_i / \mu_{crit}$ with $\mu_{crit}$
the critical density. Thus, $\mu_{X0}$ is completely fixed by the
Friedmann equation. Equation (\ref{mu-GCG}) gives \bea
   A = - w_{X0} \mu_{X0}^{1 + \alpha}.
   \label{A-w-GCG}
\eea As $A$ is determined by $w_{X0}$ and $\alpha$, in the following
we will regard $w_{X0}$ and $\alpha$ as the two free parameters to
be constrained by comparing the theoretical consequences with
observations.

The $\Lambda$CDM limit can be reached by taking $\alpha = 0$, and
identifying $A = \mu_\Lambda = \Lambda$ and $\mu_{c0} = \mu_{X0} -
A$; $\mu_c$ indicates density of the CDM. Thus, in a flat
background, from Eqs.\ (\ref{Omega_X0}) and (\ref{A-w-GCG}) we have
\bea
   w_{X0} = - {\Omega_{\Lambda 0} \over \Omega_{X0}}
       = - {\Omega_{\Lambda 0} \over \Omega_{\Lambda 0}
       + \Omega_{c0}}.
   \label{w-Lambda}
\eea We can show that even linear perturbation system (i.e.,
equations for $\delta_c$ and $v_c$) also coincides exactly with the
$\Lambda$CDM case. As a fiducial model, we use flat
$\Lambda\textrm{CDM}$ model consistent with Wilkinson Microwave
Anisotropy Probe (WMAP) 5-year data \cite{WMAP-5yr}
($\Omega_{b0}=0.0456$, $\Omega_{c0}=0.2284$, $\Omega_{\Lambda
0}=0.726$, $h=0.705$, $n_s=0.960$, $\sigma_8=0.812$,
$T_0=2.725~\textrm{K}$, $Y_\textrm{He}=0.24$, and $N_\nu=3.04$), but
without reionization history, see \cite{Park-DE-2009}. Thus we have
$w_{X0}=-0.7607$.

\begin{figure}
\begin{center}
\includegraphics[width=8.6cm]{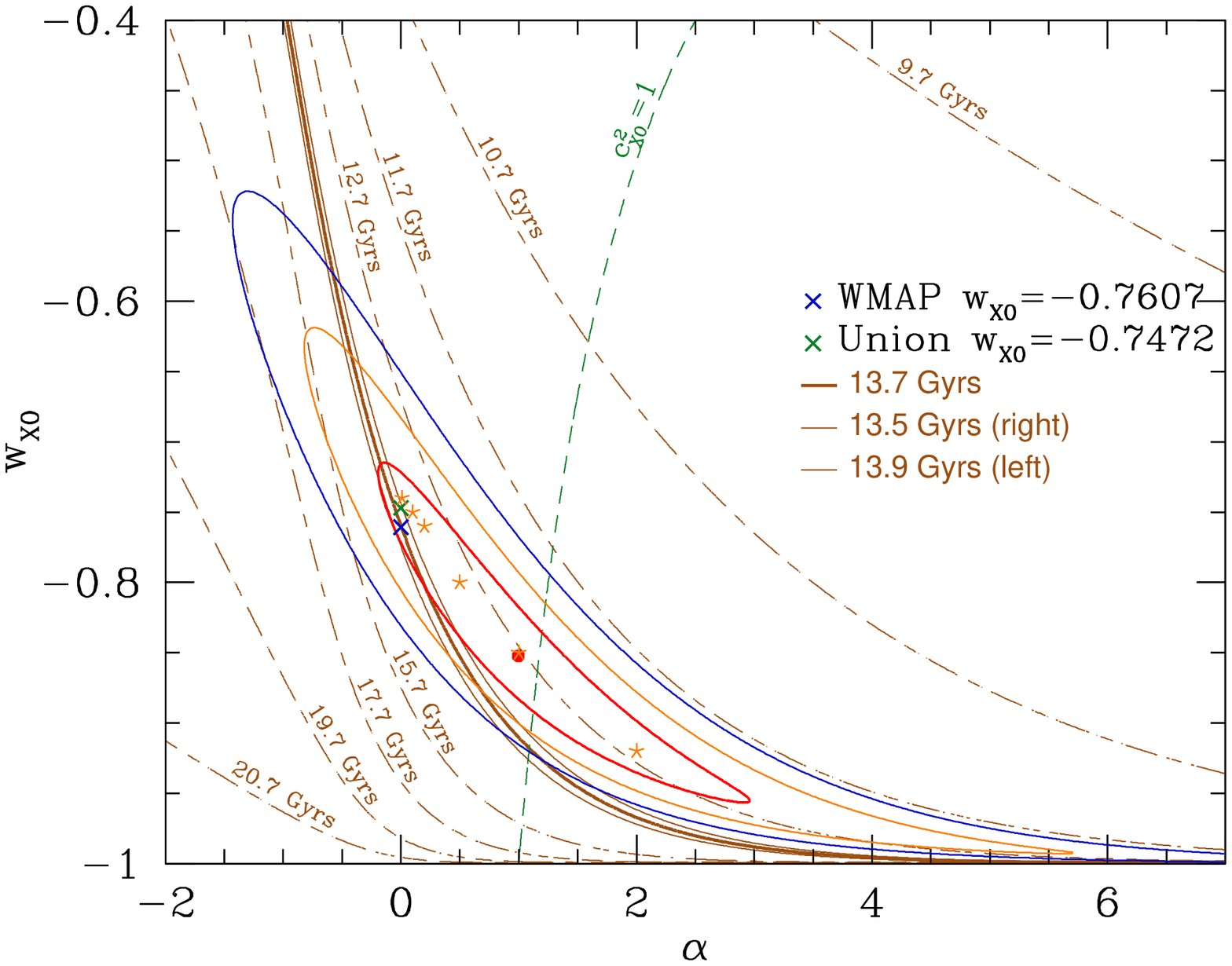} \caption{
         Likelihood (1-3$\sigma$) contours in $\alpha$-$w_{X0}$ plane
         based on the SNIa Union sample \cite{Kowalski-etal-2009}.
         The locations of $\Lambda$CDM ($\alpha = 0$) based on
         the WMAP ($\Omega_{\Lambda 0} = 0.726$) and
         Union data ($\Omega_{\Lambda 0} = 0.713$)
         are indicated as $\times$.
         The maximum likelihood point is indicated by a bullet ($\bullet$).
         Locations indicated by $\star$ are studied in
         details in Figs.\ \ref{fig:BG-evolution-1}-\ref{fig:growth-factor-1}.
         Locations near $\alpha = 0$ at
         $w_{X0} = - 0.7607$ are studied in
         details in Figs.\ \ref{fig:power-spectra-2}-\ref{fig:power-spectra-2-X}.
         We marginalized over the Hubble constant.
         Short-and-long-dashed lines are age lines in the unit of
         $(h/0.705)^{-1}$ Gyrs.
         The age lines together with the SNIa data already show that
         parameters with $\alpha >0$ are largely excluded
         by currently available observations.
         A short-dashed line shows parameters which reach $c^2_{X0} =
         1$ at the present epoch; the right side of the line has
         $c^2_{X0} > 1$ thus becoming super-luminal.
         }
\label{fig:SN-mcmc-GCG}
\end{center}
\end{figure}

\begin{figure}
\begin{center}
\includegraphics[width=8.6cm]{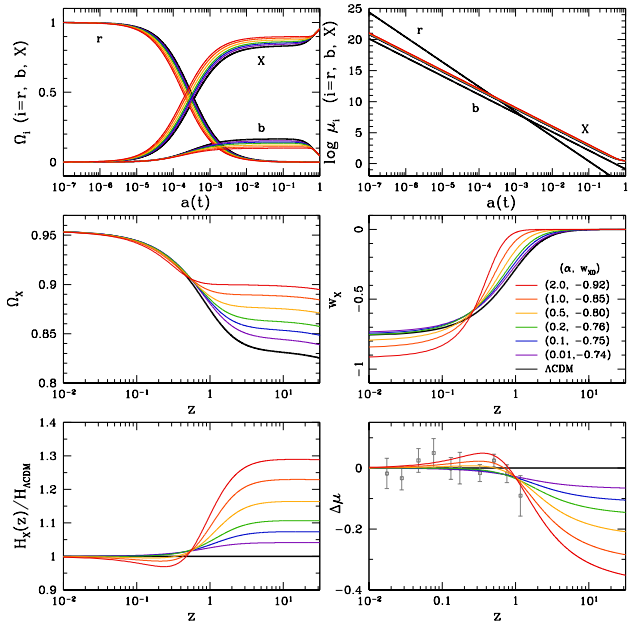}
\caption{
         Top panels: Evolution of $\Omega_i$ and $\mu_i$ as a function
         of scale factor $a(t)$ for values of $(\alpha, w_{X0})$
         indicated as $\star$ in Fig.\ \ref{fig:SN-mcmc-GCG};
         $i=\textrm{r},\textrm{b},X$ indicates radiation,
         baryon, and GCG, respectively.
         Middle and bottom panels:  Evolution of $\Omega_X$, $w_X$,
         $H_{X}(z)/H_{\Lambda\textrm{CDM}}$
         ($H_X$ is the Hubble parameter in our GCG model),
         and the relative distance modulus
         $\Delta\mu (z) = \mu_{X}(z)-\mu_{\Lambda\textrm{CDM}}(z)$
         for the same set of GCG parameters.
         In all panels, $\Lambda\textrm{CDM}$ predictions
         are shown as thick black curves.
         In the $\Delta\mu$-plot, the grey open squares with error bars
         represent the deviation of SNIa data points from the
         fiducial $\Lambda\textrm{CDM}$ model considered here.
         The binned SNIa data are based on the Union sample
         \cite{Kowalski-etal-2009}.
         }
\label{fig:BG-evolution-1}
\end{center}
\end{figure}

\begin{figure}
\begin{center}
\includegraphics[width=8.6cm]{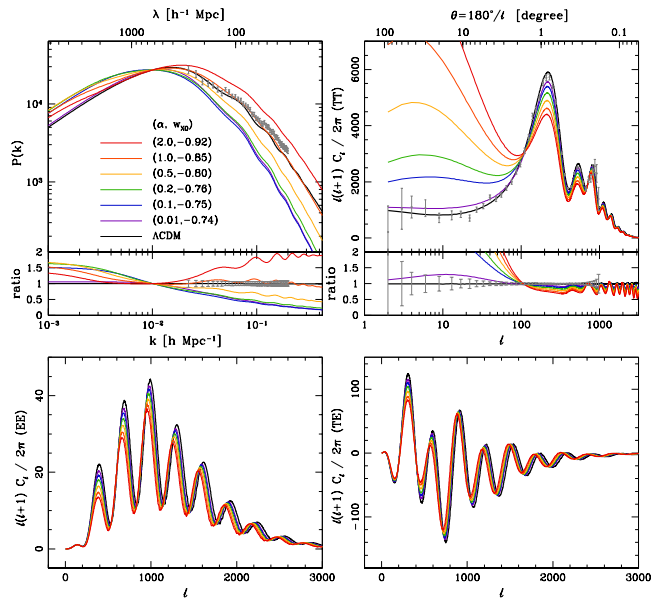}
\caption{
         The matter (baryon) power spectrum (top-left), and CMB TT (top-right), EE
         (bottom-left), TE (bottom-right) power spectra of GCG models with
         parameters used in Fig.\
         \ref{fig:BG-evolution-1}, and the same colored code.
         All calculations are made in three different gauge conditions
         (SG, UEG, and UCG), and the results in the three gauges coincide exactly.
         The matter and CMB power spectra of the $\Lambda\textrm{CDM}$ model
         have been normalized with $\sigma_8$
         and COBE spectrum, respectively.
         For comparison, all the GCG power spectra have been
         normalized with the $\Lambda\textrm{CDM}$ ones
         at $k=0.01~h$ Mpc$^{-1}$ for matter power spectrum
         and $\ell=700$ for CMB ones.
         The ratios
         of GCG powers to $\Lambda\textrm{CDM}$
         predictions are also shown in the bottom region of top
         panels.
         For matter and CMB TT power spectra,
         recent measurements from SDSS DR7 Luminous Red Galaxies (LRG)
         \cite{SDSS-DR7-LRG} and WMAP 5-year \cite{WMAP-5yr} data
         (including the cosmic variance) have been added (grey dots with
         error bars) together with fractional errors of observed spectra.
         For a correct comparison with the LRG band powers,
         the model power spectrum should include the convolution effect
         caused by LRG band power window functions and the non-linear
         clustering information.
         }
\label{fig:power-spectra-1}
\end{center}
\end{figure}

\begin{figure}
\includegraphics[width=8.6cm]{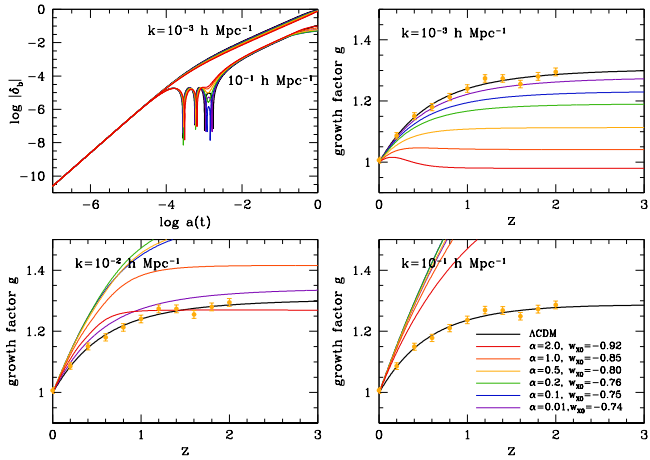} \caption{
         Evolution of baryon density perturbation (top-left),
         and the normalized perturbation growth factor $g \equiv (\delta_b/a)$
         in three different scales for the same parameters
         used in Figs.\ \ref{fig:BG-evolution-1} and \ref{fig:power-spectra-1}.
         We add $1\%$ error bar expected from future X-ray and
         weak lensing observations \cite{X-ray}.
         }
\label{fig:growth-factor-1}
\end{figure}

In Fig.\ \ref{fig:SN-mcmc-GCG} we present the likelihood contours in
$\alpha$-$w_{X0}$ plane based on the SNIa data. Also presented are
the age lines. Notice that the likelihood contour as well as the age
lines depend on $\Omega_{b0}$ in our GCG model. It is interesting to
see that, except for small region close to the $\Lambda$CDM, most of
the $1 \sigma$ region based on SNIa can be ruled out by applying age
of the universe larger than $13.5$ Gyrs. Notice that the SNIa alone
favors $\alpha = 0.9971$ and $w_{X0}=-0.8523$ (indicated by a
bullet) which will be completely excluded by the matter and the CMB
power spectra (see Fig.\ \ref{fig:power-spectra-1}) and also by the
age limit. Background evolutions for the locations indicated by
$\star$ in this Figure are presented in Fig.\
\ref{fig:BG-evolution-1}. Perturbation power spectra and the growth
factors for the same parameters are presented in Figs.\
\ref{fig:power-spectra-1} and \ref{fig:growth-factor-1}.

%
%
We consider scalar-type perturbations. For a fluid with barotropic
equation of state we naturally have \bea
   \delta p_X = c_X^2 \delta \mu_X,
   \label{pert-EOM}
\eea without the entropic perturbation. This relation is valid
without taking any gauge condition. As the anisotropic stress also
vanishes the GCG is exactly an ideal fluid. Although the fluid
definition of the GCG model is ambiguous about the ideal fluid
nature, this can be proved by the field theoretic counterpart based
on Eq.\ (\ref{action}); see Appendix A for the proof. Without taking
the temporal gauge condition the energy and the momentum
conservation equations for the GCG become \bea
   & & \dot \delta_X
       = 3 \left( w_X - c_X^2 \right) H \delta_X
   \nonumber \\
   & & \qquad
       + \left( 1 + w_X \right) \left(
       \kappa - 3 H \bar \alpha - {k \over a} v_X \right),
   \label{Ch-dot-delta-eq} \\
   & & \dot v_X = - \left( 1 - 3 c_X^2 \right) H v_X
       + {k \over a} {c_X^2 \over 1 + w_X} \delta_X
       + {k \over a} \bar \alpha,
   \label{Ch-dot-v-eq}
\eea where $c_X^2 = - \alpha w_X$ for GCG; these are Eqs.\ (A7) and
(A8) in \cite{HN-scaling-2001} with $\delta_X \equiv \delta
\mu_X/\mu_X$, $v_X$ a perturbed velocity of GCG component, $\bar
\alpha$ a perturbed metric variable and $\kappa$ a perturbed part of
the expansion scalar of the normal frame, see \cite{Bardeen-1988};
$k$ is the comoving wave number.

%
%
\section{Observational constraints}

In order to calculate the matter and CMB power spectra, and the
baryon density perturbation growth factor we solve a system composed
of baryon, radiation (handled using the Boltzmann equation or tight
coupling approximation), together with the GCG described by Eqs.\
(\ref{Ch-dot-delta-eq}) and (\ref{Ch-dot-v-eq}) representing the
dark matter and the dark energy in a unified way.  We consider a
flat background with similar parameters as our fiducial $\Lambda$CDM
model mentioned below Eq.\ (\ref{w-Lambda}) including neutrino
components. Our set of equations and the numerical methods are
presented in \cite{HN-CMB-2002}.

We solved the system in three different gauge conditions: the
synchronous gauge (SG), the uniform-expansion gauge (UEG), and the
uniform-curvature gauge (UCG); see
\cite{Park-DE-2009,Bardeen-1980,Bardeen-1988} for the description of
the gauges. The matter power spectrum and the perturbation growth
factor are for the baryonic matter perturbation in the synchronous
gauge which is a gauge-invariant concept in our situation. The CMB
temperature and polarization anisotropies are naturally gauge
invariant. The final results of these gauge-invariant variables
calculated in our three different gauge conditions should coincide;
this provides a numerical check of the calculations.

In Figs.\ \ref{fig:BG-evolution-1} and \ref{fig:power-spectra-1} we
present the background evolution, and the matter and the CMB power
spectra for several typical parameters indicated as $\star$ in Fig.\
\ref{fig:SN-mcmc-GCG}. The bottom-right panel in Fig.\
\ref{fig:BG-evolution-1} shows that all the parameters we consider
fit well with the SNIa data. However, the power spectra in Fig.\
\ref{fig:power-spectra-1} show that all the GCG models we consider
fail to fit the observed matter and CMB power spectra
simultaneously. The baryon density perturbation growth factor
presented in Fig.\ \ref{fig:growth-factor-1} also confirms this
result.


\begin{figure}
\begin{center}
\includegraphics[width=8.6cm]{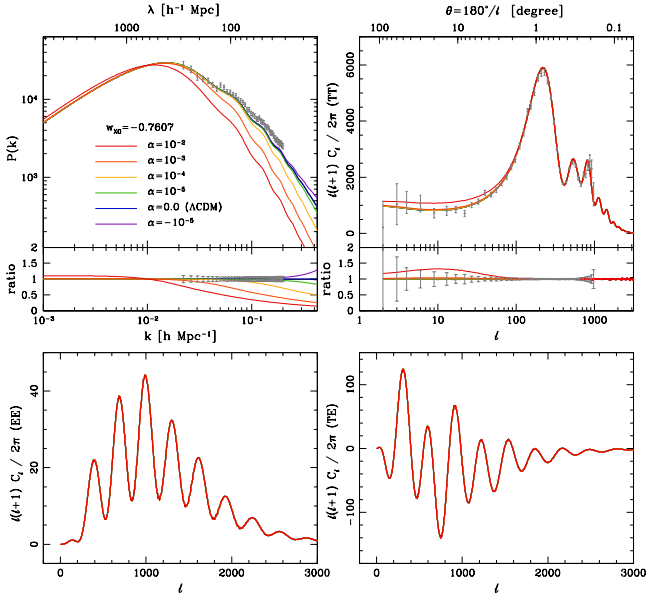}
\caption{
         The same as in Fig.\ \ref{fig:power-spectra-1}
         for values near $\alpha = 0$ (indicated in the Figure)
         with $w_{X0} = -0.7607$.
         }
\label{fig:power-spectra-2}
\end{center}
\end{figure}

\begin{figure}
\begin{center}
\includegraphics[width=8.6cm]{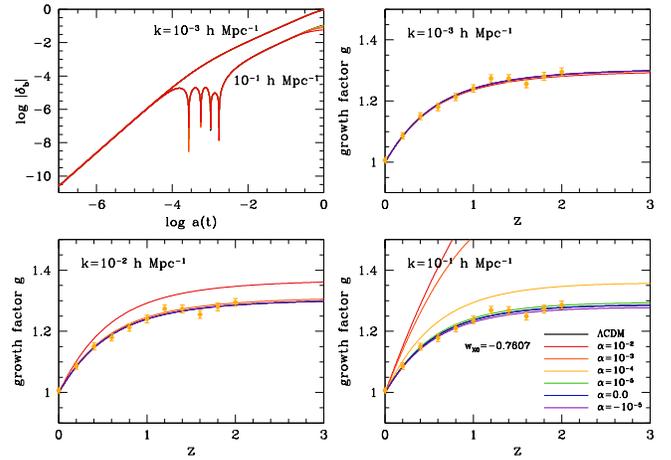}
\caption{
         The same as in Fig.\ \ref{fig:growth-factor-1}
         for the same parameters used in Fig.\
         \ref{fig:power-spectra-2}.
         }
\label{fig:growth-factor-2}
\end{center}
\end{figure}

Thus, we find that the only region in $\alpha$-$w_{X0}$ parameter
space allowed by the current observations is the location close to
$\Lambda$CDM. In order to constrain the observationally allowed
variation of $\alpha$ in that region, in Figs.\
\ref{fig:power-spectra-2} and \ref{fig:growth-factor-2} we
investigate the case of $-10^{-5} \le \alpha \le 10^{-2}$ for fixed
$w_{X0} = -0.7607$. In Fig.\ \ref{fig:power-spectra-2} we present
the matter and the CMB power spectra; we do not present the
background evolutions which are quite similar to the $\Lambda$CDM
ones in Fig.\ \ref{fig:BG-evolution-1}. For the parameters
considered the CMB power spectra are similar to the $\Lambda$CDM,
thus observationally indistinguishable for $\alpha < 10^{-2}$. The
matter power spectra in Fig.\ \ref{fig:power-spectra-2}, however,
depend more sensitively on the value of $\alpha$. For example, for
$\alpha > 10^{-4}$ the current observation can be used to
distinguish its deviation. The baryon density perturbation growth
factor presented in Fig.\ \ref{fig:growth-factor-2} also confirms
this result which shows that the deviations are particularly
significant in the small scale.

\begin{figure}
\begin{center}
\includegraphics[width=8.6cm]{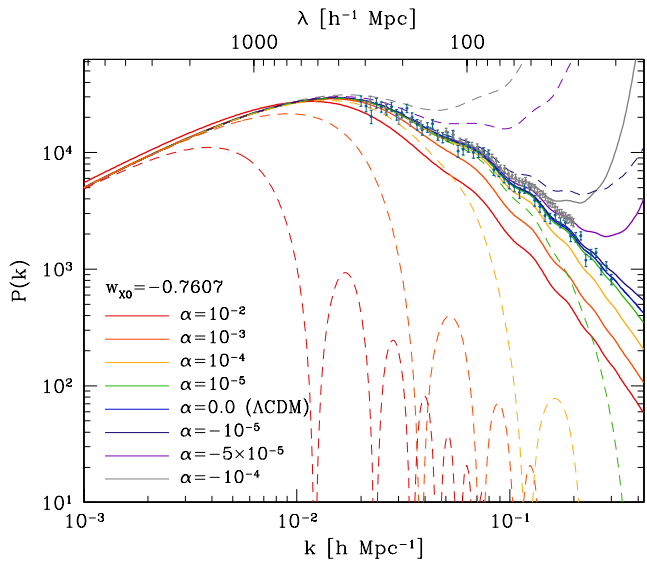}
\caption{
         The baryonic matter power spectrum in Fig.\
         \ref{fig:power-spectra-2},
         together with the GCG power spectrum plotted in dashed
         lines.
         Although the baryon power spectra are normalized at
         $k=0.01~h$ Mpc$^{-1}$, the GCG power spectra are normalized
         at $k=0.001~h$ Mpc$^{-1}$ due to heavy oscillation of the
         latter ones.
         Compared with the GCG power spectra which oscillate
         significantly, baryon power spectra relatively behave mildly.
         The grey and blue dots with error bars
         represent the SDSS DR7 LRG
         and the $\Lambda\textrm{CDM}$ mock power spectrum data,
         respectively.
         For the latter, we define 50 data points in the logarithmic interval
         between $0.02$ and $0.3~h\textrm{Mpc}^{-1}$, and perturb the linear
         $\Lambda\textrm{CDM}$ matter power spectrum at each point
         by adding a Gaussian noise with 10\% of the power spectrum
         amplitude as standard deviation. This $\Lambda$CDM-motivated mock
         data will be used to constrain our GCG model in Fig.\ \ref{fig:matter-PS-MCMC}
         and Table I.
         }
\label{fig:power-spectra-2-X}
\end{center}
\end{figure}

Based on severe oscillations and divergent behaviors in the small
scale of GCG power spectrum for nonvanishing $\alpha$, authors of
\cite{Sandvik-etal-2004} concluded that $|\alpha|$ larger than
$10^{-5}$ are excluded by the observation. This is understandable
because the sound velocity squared $c_X^2$ becomes
negative/positive, thus causing instability/oscillation for
negative/positive $\alpha$ in the small-scale limit, see Appendix B
for the analysis. However, later it has been shown that, despite the
heavy oscillation and divergence of the GCG power spectrum, the
accompanied baryon power spectrum behaves relatively well
\cite{Beca-etal-2003,Gorini-etal-2008}. In order to resolve the
issue clearly, in Fig.\ \ref{fig:power-spectra-2-X} we present the
baryon power spectra together with the GCG power spectra for the
same parameters used in Fig.\ \ref{fig:power-spectra-2}. The Figure
confirms that despite the wild oscillations and divergences of GCG
power spectra, the baryon power spectra behave much mildly. The
Figure shows, however, that although $|\alpha| \sim 10^{-5}$ are
surely acceptable, $\alpha \sim 10^{-4}$ gives deviation compared
with current observation and $\alpha \sim 10^{-3}$ could be already
excluded. For negative value of $\alpha$ we have more stringent
limit so that $\alpha < - 5.0 \times 10^{-5}$ already shows
diverging behavior in the small scale limit.

\begin{figure}
\begin{center}
\includegraphics[width=8.6cm]{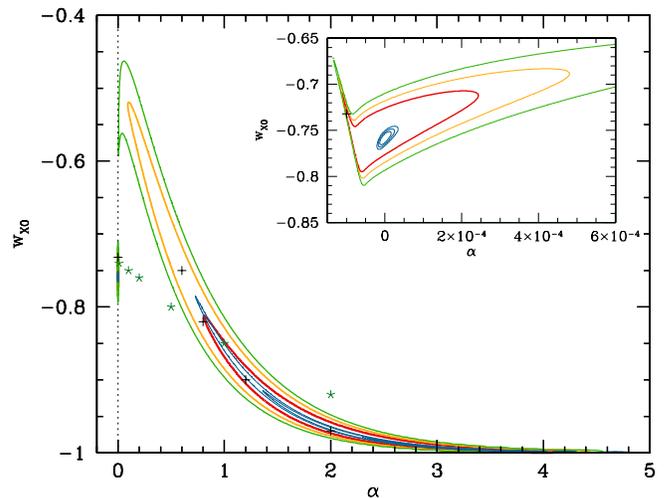}
\caption{
         Likelihood (1-3$\sigma$) contours in $\alpha$-$w_{X0}$ plane
         based on the
         matter power spectrum measurements from SDSS DR7 LRG
         \cite{SDSS-DR7-LRG} (red, yellow, green curves).
         For the method used, see Appendix C.
         Notice that besides the region around $\Lambda$CDM near
         $\alpha = 0$ (see the inner panel), the matter power spectrum
         favors another
         island with positive $\alpha$.
         This island, however, can be excluded by the CMB power
         spectra in Fig.\ \ref{fig:power-spectra-3} and the age
         limit in Fig.\ \ref{fig:SN-mcmc-GCG}.
         In the both panels, we add similar constraint
         based on $\Lambda\textrm{CDM}$-motivated mock power spectrum data
         in Fig.\ \ref{fig:power-spectra-2-X} (blue contours).
         GCG model parameters used in Figs.\
         \ref{fig:BG-evolution-1}-\ref{fig:growth-factor-1}
         are indicated by $\star$;
         power spectra for several parameters
         consistent with LRG power spectrum indicated by $+$ are presented in
         Fig.\ \ref{fig:power-spectra-3}.
         }
\label{fig:matter-PS-MCMC}
\end{center}
\end{figure}

\begin{table}
\caption{\label{tab:GCG_constraints}GCG model parameter constraints
         (68.3\% CL) from SDSS DR7 LRG and $\Lambda\textrm{CDM}$-motivated
         mock power spectrum data based on likelihood
         distribution around $\alpha=0$
         (close to $\Lambda\textrm{CDM}$ model).}
\begin{ruledtabular}
\begin{tabular}{lcc}
       & $\alpha$ & $w_{X0}$  \\[1mm]
\hline \\[-3mm]
 LRG & $-5.98_{-2.19}^{+11.3} \times 10^{-5}$
     & $-0.756_{-0.016}^{+0.023}$  \\[+1mm]
 $\Lambda\textrm{CDM}$
     & $-0.25_{-5.76}^{+5.78} \times 10^{-6}$
     & $-0.7585_{-0.0030}^{+0.0035}$  \\
\end{tabular}
\end{ruledtabular}
\end{table}

\begin{figure}
\begin{center}
\includegraphics[width=8.6cm]{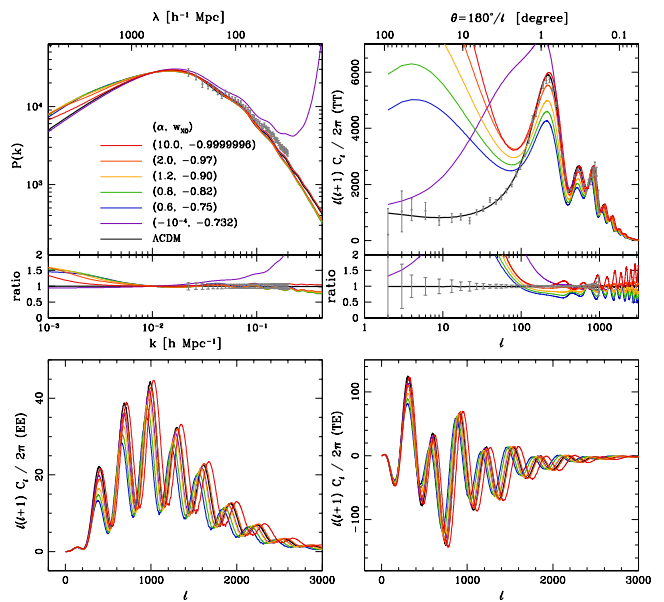}
\caption{
         The same as in Fig.\ \ref{fig:power-spectra-1}
         for parameters indicated.
         We present parameters which show similar behavior as the
         $\Lambda$CDM in the matter power spectra.
         All these models, however, show severe deviations in
         the CMB power spectra, thus are excluded.
         }
\label{fig:power-spectra-3}
\end{center}
\end{figure}

In Figs.\ \ref{fig:power-spectra-2}-\ref{fig:power-spectra-2-X} we
show that near $\alpha = 0$ (thus for GCG models close to
$\Lambda$CDM) the observed matter power spectrum can be used to
constrain tightly the allowed GCG models, whereas the CMB power
spectra are relatively similar to $\Lambda$CDM. In Fig.\
\ref{fig:matter-PS-MCMC} we show the likelihood contours of GCG
model parameters obtained from the SDSS DR7 LRG and
$\Lambda\textrm{CDM}$-motivated mock power spectrum data in Fig.\
\ref{fig:power-spectra-2-X} (see Appendix C for the method). Table
\ref{tab:GCG_constraints} lists 68.3\% confidence limits (CL) of GCG
model parameters estimated based on likelihood distribution around
the narrow region close to $\Lambda\textrm{CDM}$ model (inner panel
of Fig.\ \ref{fig:matter-PS-MCMC}). Although not favored in the age
constraint, we find another (besides the close-to-$\Lambda$CDM)
region in the $\alpha$-$w_{X0}$ parameter plane where the matter
power spectra are favored by current observations. In Fig.\
\ref{fig:power-spectra-3} we show the matter and the CMB power
spectra for several parameters within $1 \sigma$ domain in the
matter power spectrum in Fig.\ \ref{fig:matter-PS-MCMC}. Authors of
\cite{Gorini-etal-2008} showed that for large $\alpha$ with $w_{X0}$
extremely close to $-1$ the matter power spectra are compatible with
observation; this is consistent with our Fig.\
\ref{fig:matter-PS-MCMC} (note that the truncation of contour at
$\alpha\simeq 4.6$ is due to the finite bin size of $w_{X0}$ used in
the GCG model parameter search). However, our Fig.\
\ref{fig:power-spectra-3} shows that despite the observational
success in the matter power spectra, significant deviations in the
CMB power spectra are unavoidable, thus this additional domain is
excluded by the CMB observations (in addition to the age test).

%
%
\section{Discussion}
                                               \label{sec:Discussion}

We have systematically studied observational consequences of GCG
unified dark matter and dark energy model. In the background
Friedmann world model the basic requirement of minimum age of the
universe together with the SNIa luminosity-redshift data already
leaves only small parameter space close to the $\Lambda$CDM model,
see Fig.\ \ref{fig:SN-mcmc-GCG}. In the perturbation study, we have
shown that the matter power spectrum and the CMB power spectra are
mutually exclusive except for narrow region close to the
$\Lambda$CDM model, see Figs.\ \ref{fig:power-spectra-1},
\ref{fig:power-spectra-2}, and \ref{fig:power-spectra-3}. A tight
constraint on $\alpha$-parameter in that allowed region can be
obtained from the (baryonic) matter power spectrum as $-5\times
10^{-5} \le \alpha \le 10^{-4}$, see Figs.\
\ref{fig:power-spectra-2} and \ref{fig:power-spectra-2-X}. More
realistic constraint can be found in Fig.\ \ref{fig:matter-PS-MCMC}
and Table \ref{tab:GCG_constraints} as $|\alpha| \lesssim 10^{-4}$
from the SDSS DR7 LRG power spectrum and $|\alpha| \lesssim 10^{-5}$
from the $\Lambda\textrm{CDM}$ mock power spectrum data. Although
this allowed region is wider than what was concluded based on the
GCG power spectrum in \cite{Sandvik-etal-2004} (see Fig.\
\ref{fig:power-spectra-2-X}), the constraint based on the baryon
power spectrum is still severe enough so that the observationally
allowed GCG model can be regarded as extremely close to the
$\Lambda$CDM model, thus, effectively indistinguishable from the
$\Lambda$CDM model. Although named as a unified model the GCG model
has two free parameters, $\alpha$ and $w_{X0}$, whereas the
$\Lambda$CDM has only one free parameter $\Omega_{\Lambda 0}$. From
the perspective of more wider parameter space of our GCG model, it
is remarkable to notice the distinguished success and its uniqueness
of the $\Lambda$CDM model.

%
%
\subsection*{Acknowledgments}

H.N.\ was supported by grants No.\
2009-0078118 from Korea Science and Engineering Foundation (KOSEF).
J.H.\ was supported by the Korea Research Foundation (KRF) Grant
funded by the Korean Government (MOEHRD, Basic Research Promotion
Fund) (No.\ KRF-2007-313-C00322) (KRF-2008-341-C00022), and by Grant
No.\ R17-2008-001-01001-0 from KOSEF.

%
%
\section*{Appendices}

{\bf Appendix A. Tachyonic field correspondence:} Action in Eq.\
(\ref{action}) gives \bea
   & & \tilde T_{ab} = \tilde p_X \tilde g_{ab}
       - \tilde p_{X,X} \tilde \phi_{,a} \tilde \phi_{,b}.
\eea Under the energy frame, with $\tilde q_a \equiv 0$, we have
\bea
   & & \tilde \mu = - \tilde p_X + 2 \tilde p_{X,X} \tilde X, \quad
       \tilde p = \tilde p_X, \quad
       \tilde \pi_{ab} = 0,
\eea where the fluid quantities are defined in Eq.\ (2) of
\cite{GGT-1990}. From $\tilde \mu_X = - \tilde p_X + 2 \tilde
p_{X,X} \tilde X$ we can derive \bea
   & & \tilde X \propto e^{\int {2 d \tilde p_X
       \over \tilde \mu_X + \tilde p_X}}.
\eea Thus, for any barotropic fluid with $\tilde p_X = \tilde p_X
(\tilde \mu_X)$ we have a corresponding tachyonic field with $\tilde
p_X ( \tilde X)$ given by the above relation. To the background
order, we have $ c_X^2 \equiv {\dot p_X / \dot \mu_X} = {p_{X,X} /
\mu_{X,X}}$. To the perturbed order, we have $\delta p_X = p_{X,X}
\delta X$ and $\delta \mu_X = \mu_{X,X} \delta X$, thus $e_X \equiv
\delta p_X - c_X^2 \delta \mu_X = 0$. As we have $e_X = 0$ and
$\pi_{X ab} = 0$, the tachyonic field based on Eq.\ (\ref{action})
corresponds to an ideal fluid.

%
%
{\bf Appendix B. Small-scale instability for $\mbox{\boldmath
$\alpha < 0$}$:} For $\alpha < 0$ we have $c_X^2 < 0$. As the GCG is
an ideal fluid $c_X$ can be interpreted as the sound velocity, and
imaginary sound speed naturally leads to small scale instability. In
order to show the instability we take the GCG-comoving gauge which
sets $v_X \equiv 0$. Together with the Raychaudhury equation (see
Eq.\ (A14) in \cite{HN-scaling-2001}) \bea
   \dot \kappa + 2 H \kappa
       = - \left( 3 \dot H - {k^2 \over a^2} \right) \bar \alpha
       + 4 \pi G \left( \delta \mu + 3 \delta p \right),
\eea Eqs.\ (\ref{Ch-dot-delta-eq}) and (\ref{Ch-dot-v-eq}) lead to
\bea
   & & \ddot \delta_X
       + \left( 2 - 6 w_X + 3 c_X^2 \right) H \dot \delta_X
   \nonumber \\
   & & \quad
       - \left[ 3 \left( 5 w_X - 3 c_X^2 \right) H^2
       + 3 \left( w_X + c_X^2 \right) \dot H \right] \delta_X
   \nonumber \\
   & & \quad
       = {1 + w_X \over a^2 H}
       \left[ {H^2 \over a (\mu_X + p_X )}
       \left( {a^3 \mu_X \over H} \delta_X \right)^\cdot
       \right]^\cdot
   \nonumber \\
   & & \quad \quad
       + 4 \pi G \left( 1 + w \right)
       \left( 1 + 3 c_s^2 \right) \mu \delta_X
   \nonumber \\
   & & \quad
       = 4 \pi G \left( 1 + w_X \right)
       \left( \delta \mu + 3 \delta p \right)
       - c_X^2 {k^2 \over a^2} \delta_X,
\eea where $\delta \mu$ and $\delta p$ are collective (total)
perturbed energy density and perturbed pressure; $w \equiv p/\mu$
and $c_s^2 \equiv \dot p/ \dot \mu$ where $\mu$ and $p$ are
collective (total) energy density and pressure of the background
world model. We note that the above equation is generally valid in
the presence of other components (baryon, radiation, etc.) and the
background curvature. The presence of $c_X^2 k^2 \delta_X$ term in
the right-hand-side of above equation shows that for
negative/positive $c_X^2$ we have strong pressure-caused
instability/oscillation in the small scale limit where $k$ is large.

%
%
{\bf Appendix C. Likelihood estimation of GCG model parameters using
the matter power spectrum data:} Here we briefly summarize how the
likelihood contours of GCG model parameters in Fig.\
\ref{fig:matter-PS-MCMC} have been obtained from the observed and
mock power spectrum data. The power spectrum measured from SDSS DR7
LRG sample has been released recently \cite{SDSS-DR7-LRG}, with 45
band power measurements at comoving scales
$k=0.0221$--$0.199~h\textrm{Mpc}^{-1}$, corresponding band-power
window functions, and inverse covariance matrix between measurement
errors. For each GCG model of $\alpha$-$w_{X0}$, we obtain $45$ data
points by convolving the linear baryonic matter power spectrum with
the band-power window functions and normalize the point at the
largest scale ($k=0.0221~h\textrm{Mpc}^{-1}$) to that of (convolved)
fiducial $\Lambda\textrm{CDM}$ model power spectrum based on the
WMAP 5-year observation. Then we estimate the GCG model probability
distribution $\mathcal{L}\propto e^{-\chi^2/2}$ on $\alpha$-$w_{X0}$
plane by scanning $\chi^2 = {\rm \bf d}^T {\rm \bf C} ^{-1} {\rm \bf
d}$, where ${\rm \bf d}$ is a $45\times 1$ vector containing GCG
powers relative to LRG measurement and ${\rm \bf C}$ is the
$45\times 45$ covariance matrix. During the scanning, other
cosmological parameters have been fixed. By adding extremely large
noise on the diagonal components of covariance matrix corresponding
to $k > 0.1~h\textrm{Mpc}^{-1}$ scales, we have effectively excluded
LRG power spectrum information at scales where non-linear clustering
dominates. Notice that during the analysis we have completely
ignored non-linear clustering properties of the matter power
spectrum, and thus have not derived the halo power spectrum to
compare with measured LRG power spectrum. For the case of
$\Lambda\textrm{CDM}$-motivated mock power spectrum data (Fig.\
\ref{fig:matter-PS-MCMC}), similar analysis has been done but
without convolution operations.

%
%


\end{document}